\documentclass[
]{ceurart}

\usepackage{listings}
\begin{document}

\copyrightyear{2022}
\copyrightclause{Copyright for this paper by its authors.
  Use permitted under Creative Commons License Attribution 4.0
  International (CC BY 4.0).}

\conference{SC$^2$ 2021: 6th International Workshop on Satisfiability Checking and Symbolic Computation, August 19-20 2021}

\title{SC-Square: Overview to 2021}

\author{Matthew England}[%
orcid=0000-0001-5729-3420,
email=Matthew.England@coventry.ac.uk,
url=https://matthewengland.coventry.domains,
]
\address{Coventry University, Coventry, UK}

\begin{abstract}
This extended abstract was written to accompany an invited talk at the 2021 SC-Square Workshop, where the author was asked to give an overview of SC-Square progress to date.  The author first reminds the reader of the definition of SC-Square, then briefly outlines some of the history, before picking out some (personal) scientific highlights.
\end{abstract}

\begin{keywords}
symbolic computation \sep
computer algebra systems \sep 
satisfiability checking \sep
SMT solvers
\end{keywords}

\maketitle

\section{SC-Square Definition}
\label{SEC:Intro}

SC-Square, or SC$^2$, refers to the intersection of two fields of Computer Science which share that abbreviation:  Symbolic Computation and Satisfiability Checking.  The SC-Square community refers to people with an interest in \emph{both} fields.

Satisfiability Checking refers to algorithms and solvers dedicated to ascertaining whether a system of logical constraints admits a solution (allocation of values to variables which satisfies the system).  This grew out of the SAT-community and the success of SAT-solvers in answering many very large instances of the Boolean SAT Problem, despite the problem being NP-complete. It now encompasses logic problems with variables from a variety of mathematical domains.  One popular paradigm for attacking such problems is to tackle the Boolean logic separately to the constraints in the domain by viewing the atoms of the formula as Boolean variables and employing a SAT solver; then using domain specific algorithms / software to see if the domain constraints assigned to be true can be mutually satisfied.  This is often referred to as Satisfiability Modulo Theories (SMT) and the accompanying software as SMT-solvers.

Symbolic Computation refers to algorithms that perform symbolic mathematics efficiently, such as polynomial computations.  Historic achievements in symbolic computation include algorithms for  symbolic integration, polynomial factorisation, Gr\"obner bases for the effective solution of many problems concerning multivariate polynomials over algebraically-closed fields, and algorithms for addressing quantifier elimination and other problems involving a mixed system of equalities and inequalities on non-linear multivariate polynomials.  Symbolic Computation algorithms are usually implemented in Computer Algebra Systems: large software products designed for use in mathematics research and education.

Traditionally, the two communities have been largely disjoint and unaware of the achievements of one another, despite there being significant areas of overlapping interest.  In many domains of interest for SMT the theory reasoning would naturally use algorithms of symbolic computation.  In the opposite direction; the integration of SAT solvers into computer algebra systems allows for more powerful logical reasoning; and the model-driven search algorithms of satisfiability provide inspiration for whole new algorithmic approaches in computer algebra.

\section{SC-Square History}

For more information on the separate histories of the two SCs we refer the reader, for example, to Section 2 of \cite{AAB+16a}.  That paper was written to announce the start of the EU funded SC-Square Project\footnote{\url{http://www.sc-square.org/EU-CSA.html}}.  This ran from 2016$-$2018 with the aim of bridging the gap between the communities to produce individuals whom can combine the knowledge and techniques of both fields to resolve problems currently beyond the scope of either.  The project consortium consisted of a variety of EU universities, institutions and companies, and the project included a wider group of partners from all over the world.  It was formed following the invited talk of Erika \'{A}braha\'{a}m at ISSAC 2015 \cite{Abraham2015} and a 2015 Dagstuhl Seminar\footnote{\# 15471:  \url{https://www.dagstuhl.de/en/program/calendar/semhp/?semnr=15471}}.

The SC-Square Project funded new collaborations, new tool integrations, proposals on extensions to the SMT-LIB language standards, new collections of benchmarks, two summer schools in 2017\footnote{\url{http://www.sc-square.org/CSA/school/}} and 2018\footnote{\url{http://ssa-school-2018.cs.manchester.ac.uk/}}, a special issue of the Journal of Symbolic Computation (volume 100) \cite{DEGST20},  and the SC-Square Workshop Series\footnote{\url{http://www.sc-square.org/workshops.html}}.

Although the project finished in 2018, the collaborations it instigated have continued, as has the workshop series which bears its name.  There have been six editions of the workshop to date, with two more planned:
\begin{description}
\item[2016] Timisoara, Romainia (as part of SYNASC 2016).
\item[2017] Kaiserslautern, Germany (alongside ISSAC 2017).
\item[2018] Oxford, UK (as part of FLoC 2018).
\item[2019] Bern, Switzlerland (as part of SIAM AG19)
\item[2020] Paris, France (online) (alongside IJCAR 2020)
\item[2021] Texas, USA (online) (as part of SIAM AG21)
\item[2022] Haifa, Israel (planned, as part of FLoC 2022)	
\item[2023] Troms{\o} Norway (planned, alongside ISSAC 2023)
\end{description}
The workshops take place as part of, or alongside, established conferences, alternating between those in computational algebra and logic. Each year there are two chairs, one from each SC.  

\section{SC-Square Scientific Highlights}

\subsection{Non-linear Real Arithmetic}

Algorithms for working with systems of non-linear polynomials are a core topic for symbolic computation, but also an important domain for SMT where it is referred to as Non-linear Real Arithmetic (NRA).  Such algorithms are notoriously complex but also offer limitless applications, so it is no surprise that this is one area where there has been much activity\footnote{The author also acknowledges that this is his field of interest which likely led this to be forefront of his highlights!}.

Arguably the first algorithmic development in the scope of SC-Square was the NLSAT Algorithm of \cite{JdM12} which pre-dated and inspired the project.  The authors re-purposed the symbolic computation theory of cylindrical algebraic decomposition from \cite{Collins1975}, for use in their proof framework (now known as MCSAT \cite{dMJ13}) to solve satisfiability problems in NRA.

At a similar time, the SMT-RAT solver and toolbox \cite{KA18} started developing implementations of a variety of computer algebra tools for use in SMT, including cylindrical algebraic decomposition, Gr\"obner Bases, Virtual Term Substitution, and more.  This was preferable to using computer algebra systems directly as SMT theory solvers, since the algorithms needed adaption to suit the SMT requirements of efficient incrementality by constraint, backtracking and explanations for unsatisfiability.  For examples of such adaptions see e.g., \cite{KA20} for cylindrical algebraic decomposition and \cite{JLCA13} for Gr\"{o}bner bases.

The success of NLSAT and SMT-RAT inspired new algorithmic approaches.  The Conflict Driven Cylindrical Algebraic Coverings of \cite{ADEK21} gives an alternative repurposing of CAD technology for SMT, compatible with the traditional SMT proof framework.  Meanwhile the NuCAD algorithm of \cite{Brown2015} was inspired by NLSAT to use an incremental local cell construction to build a decomposition suitable for the more general quantifier elimination \cite{Brown2017} application.  This is an example of a new symbolic computation algorithm development inspired by algorithmic ideas from satisfiability checking.   

Other SC-Square work in NRA includes the combination of computer algebra system Reduce/Redlog into SMT solver VeriT \cite{FOSKT18}; the combination of computer algebra with heuristics based on interval constraint propagation and subtropical satisfiability \cite{FOSKT18}; and the Incremental Linearlization techniques of \cite{CGIRS18b}.

\subsection{Other Highlights}
	
We note a few other SC-Square highlights the author is aware of:
\begin{itemize}
\item Popular commercial computer algebra system Maple can now read to and from SMT-LIB \cite{Forrest2017} and ships with the both the Z3 and MapleSAT solvers.

\item The computer algebra system CoCoA now releases the open-source CoCoALib: a C++ Library that underpins many of its routines \cite{AB14}, more suitable for use by SMT solvers.

\item The MathCheck project has used a combination of SAT-solvers and computer algebra to make progress on a variety of combinatorics problems, enumerating new cases and verifying conjectures.  E.g., Williamson Matrices \cite{BKG20}; Golay Pairs \cite{BKHG18}; Good Matrices \cite{BDKG19}.

\item Algebraic techniques were key for the circuit verification work \cite{Kaufmann2021}, and in combination with SAT-solvers \cite{KBK19}.

\item The Boolean SAT problem has also benefited from symbolic computation via Boolean Gr\"{o}bner Bases and parallel computation on both the conjunctive and algebraic normal forms of a problem \cite{HK20}. 

\item An emerging direction in both SCs is the production of proofs and some initial ideas here have emerged from SC-Square: \cite{Jebelean2018}, \cite{ADEKT20}.

\item The SC-Square workshops have been home to application descriptions that are new to both SCs, including in the fields of Economics \cite{MBDET18}, Dynamic Geometry \cite{VK20}, and knot theory \cite{LV17}, \cite{MP18}.  

\end{itemize}

\section{SC-Square Future}

The past successes of SC-Square warrant a promising future for the community.  The workshop series will continue in 2022 and 2023 at the least.  There will also be a new Dagstuhl Seminar on the topic\footnote{\# 22072:  \url{https://www.dagstuhl.de/en/program/calendar/semhp/?semnr=22072}}.  However, it is still the case that the bulk of both communities are working independently from each other, and greater integration would surely bring further successes.

\subsubsection*{Acknowledgements}  

The author is supported by the EPSRC project, \emph{Pushing Back the Doubly-Exponential Wall of Cylindrical Algebraic Decomposition} (EP/T015748/1).  Thanks to the reviewer for suggestions that improved the text.


\end{document}